\documentclass[10pt,conference,letter]{IEEEtran}
\usepackage{graphics}
\usepackage{bm}
\usepackage{amsmath}
\usepackage{epstopdf}
\usepackage{graphicx}
\usepackage{amsmath}
\usepackage{amsfonts}
\usepackage{color}
\usepackage[]{algorithm2e}
 \usepackage{tikz}
\usepackage[utf8]{inputenc}
\usepackage[english]{babel}

\usepackage{amsmath,amssymb,amsfonts,amsbsy}
\usepackage{color}
\usepackage{siunitx}
\usepackage{graphicx}
\usepackage{booktabs}

\usepackage{graphics}
\usepackage{bm}
\usepackage{amsmath}
\usepackage{epstopdf}
\usepackage{graphicx}
\usepackage{amsmath}
\usepackage{amsfonts}
\usepackage{color}
\usepackage[]{algorithm2e}
\let\MYoriglatexcaption\caption
\renewcommand{\caption}[2][\relax]{\MYoriglatexcaption[#2]{#2}}

\renewcommand{\eqref}[1]{(\ref{#1})}

\begin{document} 
	
\title{Performance Limits of Energy Detection Systems with Massive Receiver Arrays}

\author{\begin{tabular}{c}
Lishuai~Jing$^\dagger$, Zoran~Utkovski$^\#$$^*$, Elisabeth~de~Carvalho$^\dagger$, and Petar~Popovski$^\dagger$\\
$^\dagger$Antennas, Propagation and Radio Networking (APNet) Section, Electronic Systems, Aalborg University, Denmark\\
$^\#$ Faculty of Computer Science, University Goce Del\v{c}ev \v{S}tip, R. Macedonia \\
$^*$ Laboratory for Complex Systems and Networks, Macedonian Academy of Sciences and Arts\\
Email: \{lji,edc,petarp\} @es.aau.dk, zoran.utkovski@ugd.edu.mk\\
\end{tabular}}

%


\maketitle

\begin{abstract}
Energy detection (ED) is an attractive technique for symbol detection at receivers equipped with a large number of antennas, for example in millimeter wave communication systems. This paper investigates the performance bounds of ED with pulse amplitude modulation (PAM) in large antenna arrays under single stream transmission and fast fading assumptions. The analysis leverages information-theoretic tools and semi-numerical approach to provide bounds on the information rate, which are shown to be tight in the low and high signal-to-noise ratio (SNR) regimes, respectively. For a fixed constellation size, the impact of the number of antennas and SNR on the achievable information rate is investigated. Based on the results, heuristics are provided for the choice of the cardinality of the adaptive modulation scheme as a function of the SNR and the number of antennas.
\end{abstract}
\section{Introduction}
\label{Dsection:introduction}
The introduction of large (massive) antenna arrays at wireless transceivers may bring significant enhancement in communication throughput and reliability \cite{T.L.Marzetta2010}. Such systems not only maintain the benefits introduced by traditional multiple-input-multiple-output (MIMO) systems, but also allow for the employment of linear processing techniques which reduce the computation complexity and potentially increase the energy efficiency 
\cite{F.RusekD.PerssonKiongLauBuonE.G.LarssonT.L.MarzettaO.EdforsandF.Tufvesson2013}, \cite{Lu2014}. The requirement on sufficiently accurate channel state information (CSI) to facilitate coherent processing at multiple antennas appears to be crucial for the performance of massive MIMO systems  \cite{F.RusekD.PerssonKiongLauBuonE.G.LarssonT.L.MarzettaO.EdforsandF.Tufvesson2013}. In addition, the issue of CSI acquisition is the main reason why cellular massive MIMO systems (at least in the frequency bands below $6$ GHz) are restricted to operate in TDD mode, instead of the more (industry-)friendly FDD mode. 

While the application of millimeter wave (mm-wave) frequencies in communication systems enables both base stations and small terminals to carry a large number of antennas, the deployment of large antenna arrays operating in this portion of the frequency spectrum comes with certain challenges. For example, for the same mobility, in mm-wave systems the coherence time will be an order of magnitude shorter due to higher
Doppler spread, which may reduce both spatial multiplexing
and CSI acquisition capabilities. In addition, hardware implementation may be quite different from what has been considered in the Massive MIMO literature. 

These observations motivate the use of alternative communication and signal processing techniques which are robust to CSI imperfections. A prominent example are the so-called ``non-coherent" communication/processing techniques which do not rely on (explicit) CSI acquisition/estimation. 
Energy detection (ED) (in combination with an appropriate transmission/modulation technique) is an example of a noncoherent signal processing technique which does not rely (at least not explicitly) on CSI acquisition. Due to its phase-independent property and low implementation complexity, ED lends itself as tangible solution for mm-wave signal processing. As result, PAM-ED is selected for wide-band mm-wave short range communication standards, such as Ecma-$387$ and IEEE $802.15.3$ \cite{Baykas2011}. When employed at a large antenna array receiver, ED allows for symbol detection performed without the knowledge of the instantaneous channel coefficients. In fact, with a large antenna array, ED may even operate without explicit knowledge of the channel statistics, as signal squaring and averaging performed over the excessive number of receive antennas provides a sample mean-based estimate of the channel energy. Meanwhile, due to noise hardening, additive noise contribution asymptotically approaches to a deterministic term. 

For these reasons, ED-based systems intended for mm-wave applications have been in the focus of recent research efforts. In \cite{Chowdhury2014a}, the authors show that energy detection achieves comparable symbol-error-rate (SER) against coherent detection in which the CSI is assumed to be perfectly known. In \cite{Martinez2014}, threshold values are optimized to reduce the SER based on Gaussian approximations of the energy detector output. Using SER as an optimization criteria, asymptotically optimal signal energy constellation design are proposed in \cite{Manolakos2015}. While SER is a sensible measure for the performance of ED systems equipped with massive antennas, performance limits in terms of achievable information rates are still not well-investigated.         

In this contribution we derive performance bounds of PAM-ED systems with large antenna arrays under single stream transmission. We use information-theoretic tools and semi-numerical analysis to evaluate an upper and lower bound on the mutual information. The results, which are valid for the whole SNR range, illustrate that the gap between the bounds vanishes in the low-SNR and in the high-SNR regime. For fixed constellation size, we analyze the impact of the number of antennas and SNR on the achievable rate. In addition, we provide heuristics for the selection of adaptive modulation techniques which appropriately match the cardinality of the constellation to the SNR and the number of antenna elements. Finally, we compare the obtained bounds against the capacity of the memoryless fading SIMO channel studied in \cite{MayukhSengupta2000}, which serves as benchmark for the performance of any SIMO scheme (including ED) which operates under fast, memoryless fading. 


\section{System Model}
\label{Dsection:SignalModel}
ED can be seen as a special form of processing where the received signal at each antenna is passed through a squaring device followed by integration and sampling. As ED is restricted to nonnegative modulation alphabets, we assume that the transmitted signal is $\sqrt{\boldsymbol{x}}$ (instead of $\boldsymbol{x}$). We focus on PAM such that $\boldsymbol{x}$ is selected from the energy constellation set $ \mathcal{X} = \{ 0, \epsilon_1, \epsilon_2, \dots, \epsilon_{P-1}\}$,  where $\epsilon_p= c p >0$, and $c$ denotes a normalization constant. Equipped with this notation, the output of the energy detector is written as 
\begin{align}\label{EDout}
\boldsymbol{z} & = \frac{1}{M} \sum^{M}_{i = 1} \left| \boldsymbol{h}_i \sqrt{\boldsymbol{x}} + \boldsymbol{n}_i \right|^2\\
&= \underbrace{\frac{1}{M} \sum^{M}_{i = 1} | \boldsymbol{h}_i |^2}_{\boldsymbol{\varsigma}_{h}} \boldsymbol{x}  + \underbrace{\frac{1}{M} \sum^{M}_{i = 1} |\boldsymbol{n}_i |^2}_{\boldsymbol{\varsigma}_{n}} + \underbrace{\frac{2}{M} \sum^{M}_{i = 1} \Re \left(\boldsymbol{h}_i \boldsymbol{n}^{*}_i\right)}_{\boldsymbol{w}} \sqrt{\boldsymbol{x}}.\nonumber
\end{align}
\noindent We assume that the channel coefficients are i.i.d complex Gaussian, $\boldsymbol{h}_i\sim CN (0,\sigma^2_h)$. Hence, the sample mean $\boldsymbol{\varsigma}_{h}\doteq\frac{1}{M} \sum^{M}_{i = 1} | \boldsymbol{h}_i |^2$ has the same distribution as $\frac{\sigma^2_h}{2}\bar{\boldsymbol{\varsigma}}_{h}$, where $\bar{\boldsymbol{\varsigma}}_{h}$ is Gamma distributed with shape $M$ and scale $2/M$, $\bar{\boldsymbol{\varsigma}}_{h}\sim \Gamma(M, 2/M)$ \footnote{For $\boldsymbol{h}_i\sim CN (0,\sigma^2_h)$, $\frac{2}{\sigma^2_h}| \boldsymbol{h}_i |^2$ is chi-square distributed with $2$ degrees of freedom.}. The noise coefficients are i.i.d Gaussian with zero mean and variance $\sigma^2_n$, $\boldsymbol{n}_i\sim CN (0,\sigma^2_n)$. As result, the sample mean $\boldsymbol{\varsigma}_{n}\doteq\frac{1}{M} \sum^{M}_{i = 1} | \boldsymbol{n}_i |^2$ has the same distribution as $\frac{\sigma^2_n}{2}\bar{\boldsymbol{\varsigma}}_{n}$, where $\bar{\boldsymbol{\varsigma}}_{n}$ is Gamma distributed with shape $M$ and scale $2/M$, $\bar{\boldsymbol{\varsigma}}_{n}\sim \Gamma(M, 2/M)$. 
For $M$ large, as result of the Central Limit Theorem (CLT), one can invoke a Gaussian approximation for the variables $\boldsymbol{\varsigma}_{h}$ and $\boldsymbol{\varsigma}_{n}$: 
\begin{align} \label{GauApp}
\boldsymbol{\varsigma}_{h}\sim N\left(\sigma^2_h,\sigma^4_h/M\right);\:
\boldsymbol{\varsigma}_{n}\sim N\left(\sigma^2_n,\sigma^4_n/M\right).
\end{align}
\noindent Similar argumentation holds for the statistics of $\boldsymbol{w}$
\begin{align}
\boldsymbol{w}&\doteq\frac{2}{M}\sum_{i=1}^M \Re(\boldsymbol{h}_i(k)\boldsymbol{n}^*_i(k))\sim N(0,2\sigma^2_h\sigma^2_n/M).
\end{align}
We note that the Gaussian approximations are well justified when $M$ is sufficiently large, which is a preferable operating region for massive MIMO systems. An important consequence of the detection process is that, as long as requirements for the applicability of CLT are fulfilled, signal processing based on ED is robust both with respect to the channel and noise statistics. 
Indeed, one may observe that as $M \rightarrow +\infty$, the (random) \textit{channel variables} $\boldsymbol{\varsigma}_{h}$, $\boldsymbol{\varsigma}_{n}$ and $\boldsymbol{w}$ become more deterministic, as $\boldsymbol{\varsigma}_{h} \rightarrow \sigma^2_h$, $\boldsymbol{\varsigma}_{n} \rightarrow \sigma^2_n$ and $\boldsymbol{w}\rightarrow 0$. 
This indicates that, for large antenna array systems, the first and second moments of the channel and noise may be sufficient to characterize the output of the energy detector. Due to this asymptotic behaviour, reliable detection of the transmitted symbol $\sqrt{\boldsymbol{x}}$ may be performed using the statistics of $\boldsymbol{\varsigma}_{h}, \boldsymbol{\varsigma}_{n}$ and $\boldsymbol{w}$. With this implementation, estimation of the individual channel coefficients or the individual channel amplitudes (as it is the case with conventional energy detection with a small number of antennas) is not required, which is a remarkable feature of ED schemes with massive receiver arrays. 
\section{Performance limits of ED}
An exact information-theoretic characterization of the channel described by (\ref{EDout}) is difficult in general, mainly due to the technicalities in the evaluation of the involved differential entropies under the assumption of no a-priori knowledge of the channel variables. 
In the following we evaluate lower and upper bounds on the mutual information $I(\boldsymbol{x};\boldsymbol{z})$ by relying on (simple) bounding techniques and semi-numerical analysis.
\vspace{-5pt}
\subsection{Lower Bounds}
\label{Dsection:LB}
\vspace{-3pt}
\subsubsection{Conditioning on $\boldsymbol{\varsigma}_{h}$}  
The mutual information between $\boldsymbol{x}$ and $\boldsymbol{z}$ may be lower-bounded as
\begin{align}
I(\boldsymbol{x};\boldsymbol{z})&=h(\boldsymbol{z})-h(\boldsymbol{z}|\boldsymbol{x})\geq h(\boldsymbol{z}|\boldsymbol{\varsigma}_{h})-h(\boldsymbol{z}|\boldsymbol{x}),
\label{eq:Lower_Bound_1}
\end{align}
where $h(\cdot)$ denotes differential entropy and the inequality follows from the fact that conditioning on $\boldsymbol{\varsigma}_{h}$ does not increase the entropy of $\boldsymbol{z}$. 
To compute $h(\boldsymbol{z}|\boldsymbol{x})$, we write
\begin{equation}
h(\boldsymbol{z}|\boldsymbol{x})=\mathbb{E}_{\boldsymbol{x}} \left[h(\boldsymbol{z}|\boldsymbol{x}=x)\right],
\end{equation}
where $\mathbb{E}_{\boldsymbol{x}}$ denotes that the expectation is taken over the random variable $\boldsymbol{x}$. Note that 
the pdf $f_{\boldsymbol{z}}(z|\boldsymbol{x}=x)$ is Gaussian with first moment $\sigma_{h}^2 x+\sigma_{n}^2$ and second moment $\frac{\sigma_{h}^4 x^2+ \sigma_{n}^4 +2\sigma^2_h\sigma^2_n x}{M}$.
Hence, $h(\boldsymbol{z}|\boldsymbol{x})$ reads 
\begin{equation}
h(\boldsymbol{z}|\boldsymbol{x})=\frac{1}{2}\mathbb{E}_{\boldsymbol{x}}\left[\ln \left( \frac{2\pi e(\sigma_{h}^4 x^2+ \sigma_{n}^4 +2\sigma^2_h\sigma^2_n x)}{M}\right)\right].
\label{eq:Conditional_Entropy}
\end{equation}
To derive $h(\boldsymbol{z}|\boldsymbol{\varsigma}_{h})$, we proceed by writing
\begin{align}
&h(\boldsymbol{z}|\boldsymbol{\varsigma}_{h})=\mathbb{E}_{\boldsymbol{{\varsigma}}_{h}} [h(\boldsymbol{z}|\boldsymbol{\varsigma}_{h}=\varsigma_{h})]\nonumber\\
&=-\int f_{\boldsymbol{\varsigma}_{h}}(\varsigma_{h}) \int f_{\boldsymbol{z}}(z|\boldsymbol{\varsigma}_{h}=\varsigma_{h})\log_2 f_{\boldsymbol{z}}(z|\boldsymbol{\varsigma}_{h}=\varsigma_{h})\mathrm{d}z\mathrm{d}\varsigma_{h},
\label{eq:Entropy_z_h} 
\end{align}
where $f_{\boldsymbol{\varsigma}_{h}}(\varsigma_{h})$ is a Gaussian pdf, see \eqref{GauApp}. By the law of total probability, $f_{\boldsymbol{z}}(z|\boldsymbol{\varsigma}_{h}=\varsigma_{h})$ can be computed as
\begin{align}
f_{\boldsymbol{z}}(z|\boldsymbol{\varsigma}_{h}=\varsigma_{h})
& =\sum_{x}P_{\boldsymbol{x}}(x)\cdot f_{\boldsymbol{z}}(z|\boldsymbol{\varsigma}_{h}=\varsigma_{h},\boldsymbol{x}=x),
\label{GauMix}
\end{align}
where $P_{\boldsymbol{x}}(x)$ is the probability mass function of $\boldsymbol{x}$.
Given $\varsigma_{h}$ and $x$, $\boldsymbol{z}$ is Gaussian with first moment $\varsigma_{h} x+\sigma^2_n$ and second moment $\frac{\sigma^4_n+2\sigma^2_h\sigma^2_n x}{M}$. As \eqref{GauMix} is a Gaussian mixture, we rely on Monte Carlo averaging to obtain an accurate approximation of $h(\boldsymbol{z}|\boldsymbol{\varsigma}_{h}=\varsigma_{h})$. Similarly, Monte Carlo averaging can be applied to evaluate the outer integral in \eqref{eq:Entropy_z_h}. 
\vspace{-3pt}
\subsubsection{Conditioning on $\boldsymbol{w}$} 
Alternatively, a lower bound on $I(\boldsymbol{x};\boldsymbol{z})$ may be obtained by conditioning on $\boldsymbol{w}$:
\begin{align}
I(\boldsymbol{x};\boldsymbol{z})&\geq h(\boldsymbol{z}|\boldsymbol{w})-h(\boldsymbol{z}|\boldsymbol{x}).
\label{eq:Lower_Bound_2}
\end{align}
The difference with respect to \eqref{eq:Lower_Bound_1} is in the first term on the right hand side of (\ref{eq:Lower_Bound_2}). In fact, the motivation to evaluate these lower bounds 
is that expectation that (\ref{eq:Lower_Bound_1}) and (\ref{eq:Lower_Bound_2}) may represent an accurate low-SNR, respectively high-SNR approximation of $I(\boldsymbol{x};\boldsymbol{z})$. This conclusion is motivated by the observation that in the low-SNR regime, the entropy of $\boldsymbol{z}$ is dominated by the term $\boldsymbol{w}\sqrt x$ and $h(\boldsymbol{z})\approx h(\boldsymbol{z}|\boldsymbol{\varsigma}_{h})$. Similarly, in the high-SNR regime, $\boldsymbol{\varsigma}_{h} x $ dominates the entropy of $\boldsymbol{z}$ and thus $h(\boldsymbol{z})\approx h(\boldsymbol{z}|\boldsymbol{w})$.

\noindent The term $h(\boldsymbol{z}|\boldsymbol{w})$ reads
\begin{align}
&h(\boldsymbol{z}|\boldsymbol{w})=\mathbb{E}_{\boldsymbol{w}} \left[h(\boldsymbol{z}|\boldsymbol{w}=w) \right]\nonumber\\
&=-\int f_{\boldsymbol{w}}(w) \int  f_{\boldsymbol{z}}(z|\boldsymbol{w}=w)\ln f_{\boldsymbol{z}}(z|\boldsymbol{w}=w)\mathrm{d}z\mathrm{d}w,
\label{eq:Entropy_z_w} 
\end{align}
with the pdf of the Gaussian mixture $f_{\boldsymbol{z}}(z|\boldsymbol{w}=w)$ given by
\vspace{-5pt}
\begin{equation}
f_{\boldsymbol{z}}(z|\boldsymbol{w}=w)=\sum_{x}P_{\boldsymbol{x}}(x)\cdot f_{\boldsymbol{z}}(z|\boldsymbol{w}=w,\boldsymbol{x}=x),
\vspace{-5pt}
\end{equation} 
where $f_{\boldsymbol{z}}(z|\boldsymbol{w}=w,\boldsymbol{x}=x)$ is Gaussian with first moment $\sigma^2_h x +w\sqrt x  +\sigma^2_n$ and second moment $\frac{\sigma^4_h x^2 + \sigma^4_n}{M}$. 



\vspace{-5pt}
\subsection{Upper Bounds}
\label{Dsection:UB}
Upper bounds on $I(\boldsymbol{x};\boldsymbol{z})$ may be computed if the instantaneous realization of $\boldsymbol{\varsigma}_{h}$ or $\boldsymbol{w}$ is known. This may be seen as a form of \textit{coherent}, i.e. \textit{genie-aided} detection.
\vspace{-3pt}
\subsubsection{An upper bound based on the knowledge of  $\boldsymbol{\varsigma}_{h}$} \label{UBh}
We assume that the (instantaneous) realization of $\boldsymbol{\varsigma}_{h}$ is known to the receiver (meaning that the estimate of $\boldsymbol{\varsigma}^2_h$ is perfect). In this case, it holds
\begin{align}
I(\boldsymbol{x};\boldsymbol{z})& \leq I(\boldsymbol{x};\boldsymbol{z}|\boldsymbol{\varsigma}_{h})=h(\boldsymbol{z}|\boldsymbol{\varsigma}_{h})-h(\boldsymbol{z}|\boldsymbol{\varsigma}_{h},\boldsymbol{x}).
\label{eq:Upper_Bound_1}
\end{align}
As $h(\boldsymbol{z}|\boldsymbol{\varsigma}_{h})$ has already been evaluated in (\ref{eq:Entropy_z_h}), it remains to evaluate
\begin{equation}
h(\boldsymbol{z}|\boldsymbol{\varsigma}_{h},\boldsymbol{x})=\sum_{x}P_{\boldsymbol{x}}(x)\int f_{\boldsymbol{\varsigma}_{h}}(\varsigma_{h})h(\boldsymbol{z}|\boldsymbol{\varsigma}_{h}=\varsigma_{h},\boldsymbol{x}=x)\:\mathrm{d}\varsigma_{h}.
\label{eq:Conditional_h_z_h}
\end{equation}
Given $\varsigma_{h}$ and $x$, $\boldsymbol{z}$ is Gaussian with mean $\varsigma_{h}x  + \sigma^2_n$ and variance $\frac{2\sigma^2_h\sigma^2_n x + \sigma^4_n}{M}$. Thus, we obtain
\begin{equation}
h(\boldsymbol{z}|\boldsymbol{\varsigma}_{h}=\varsigma_{h},\boldsymbol{x}=x)=\frac{1}{2}\left[\ln \left( \frac{2\pi e(2\sigma^2_h\sigma^2_n x + \sigma^4_n)}{M}\right)\right]. \notag
\end{equation}
Due to the independence of $h(\boldsymbol{z}|\boldsymbol{\varsigma}_{h}=\varsigma_{h},\boldsymbol{x}=x)$ on $\varsigma_{h}$, (\ref{eq:Conditional_h_z_h}) simplifies to
\begin{equation} \label{eq:Conditional_h_z_h1}
h(\boldsymbol{z}|\boldsymbol{\varsigma}_{h},\boldsymbol{x})=\frac{1}{2}\sum_{x}P_{\boldsymbol{x}}(x)\cdot\ln \left( \frac{2\pi e(2\sigma^2_h\sigma^2_n x + \sigma^4_n)}{M}\right).
\end{equation}
Compared to \eqref{eq:Conditional_Entropy}, the difference is that the channel energy dependent term is omitted, which reduces the entropy of $\boldsymbol{z}$. The results in Section \ref{Dsection:Results} suggest that \eqref{eq:Upper_Bound_1} and \eqref{eq:Lower_Bound_1} coincide in the low-SNR regime due to the dominance of $\sigma^4_n$. 
\subsubsection{An upper based on the knowledge of $\boldsymbol{w}$} \label{UBw}
Alternatively, an upper bound on $I(\boldsymbol{x};\boldsymbol{z})$ may be derived by assuming that the receiver knows the (instantaneous) realization of $\boldsymbol{w}$
\begin{align}
I(\boldsymbol{x};\boldsymbol{z})& \leq I(\boldsymbol{x};\boldsymbol{z}|\boldsymbol{w}) =h(\boldsymbol{z}|\boldsymbol{w})-h(\boldsymbol{z}|\boldsymbol{w},\boldsymbol{x}).
\label{eq:Upper_Bound_2}
\end{align}

\vspace{-10pt}
Having already evaluated $h(\boldsymbol{z}|\boldsymbol{w})$, it remains to compute
\begin{equation}
h(\boldsymbol{z}|\boldsymbol{w},\boldsymbol{x})=\sum_{x}P_{\boldsymbol{x}}(x)\int f_{\boldsymbol{w}}(w)h(\boldsymbol{z}|\boldsymbol{w}=w,\boldsymbol{x}=x)\:\mathrm{d}w. \notag
\label{eq:Conditional_h_z_w}
\end{equation}
Given $w$ and $x$, $\boldsymbol{z}$ is Gaussian with mean $\sigma^2_h x + w\sqrt{x} + \sigma^2_n$ and variance $\frac{\sigma^4_h x^2 + \sigma^4_n}{M}$. It is straightforward to show that
\begin{equation}\label{IXZw}
h(\boldsymbol{z}|\boldsymbol{w},\boldsymbol{x})=\frac{1}{2}\sum_{x}P_{\boldsymbol{x}}(x)\cdot\ln \left( \frac{2\pi e(\sigma^4_h x^2 + \sigma^4_n)}{M}\right).
\end{equation}
In the high SNR regime, i.e. with vanishing noise, conditioning on $\boldsymbol{w}$ has negligible impact on the entropy of $\boldsymbol{z}$. Therefore, it is expected that the lower bound evaluated in \eqref{eq:Lower_Bound_1} and the upper bound evaluated in \eqref{eq:Upper_Bound_2} will be close to each other. 
\vspace{-5pt}
\subsection{SIMO Capacity as a Performance Benchmark}
Although ED is robust to the channel statistics, we choose to compare its performance (in terms of achievable rate) with the capacity of the SIMO channel with memoryless (i.e. fast) Rayleigh fading. This model is motivated by the fact that, in the presented form, ED relies on symbol-by-symbol detection and does not exploit channel correlation in time (if any). In addition, the short coherence time in mm-wave systems justifies a fading model (e.g. Gauss-Markov) with small temporal correlation. When $M\gg 1$, a closed-form expression for the capacity of the memoryless Rayleigh fading SIMO channel with both peak and average power constraint has been derived in \cite{MayukhSengupta2000}. For average power constraint and signal-to-noise ratio $\rho$, the capacity reads 
\vspace{-5pt}
\begin{equation}
C(\rho)=\frac{1}{2}\log_2 \left(\frac{M}{2\pi}\right)+\log (a)+\frac{\rho}{a(1+\rho)},
\label{eq:SIMO_capacity_memoryless}
\end{equation}
where $a$ is a normalization constant related to the capacity-achieving distribution provided in  \cite{MayukhSengupta2000}). The expression (\ref{eq:SIMO_capacity_memoryless}) reveals a logarithmic scaling behaviour as function of the number of antennas $M$ and is used as comparison benchmark for the achievable information rate of PAM-ED.    
\vspace{-3mm}
\section{Performance Evaluation}
\label{Dsection:Results}
We evaluate the proposed bounds on the achievable information rate for the PAM-ED schemes. We remark that, for discrete input constellations, the achievable information rate is (trivially) upper-bounded by the cardinality of the modulation scheme. 
The simulation setting and notation are reported in Table \ref{Dtab:para}. 
\begin{table}[h]
	\caption{Simulation Settings and Notations}
	\centering
	\begin{tabular}{l} \toprule 
		\quad $\sigma^2_h$ = 1, SNR = $ \frac{E[\boldsymbol{x}]}{\sigma^{2}_n} $, $P_X(x):$ Uniform distribution.\\
		\midrule
		\textbf{Notations}: \\
		\quad LB:H (W) \quad Lower bound conditioned on $\boldsymbol{\varsigma}_{h}$ ($\boldsymbol{w}$).  \\
		\quad UB:H (W) \quad Upper bound conditioned on $\boldsymbol{\varsigma}_{h}$ ($\boldsymbol{w}$). \\
		\quad LB: $\doteq \max (\text{LB:H}, \text{LB:W})$, UB: $\doteq \min (\text{UB:H}, \text{UB:W})$ \\
		\bottomrule
	\end{tabular}
	\label{Dtab:para} 
\end{table} 
We plot the ``composite" curve which is obtained by taking the maximum of the two lower bounds for each value of the SNR. Similarly, from the two upper bounds, we select the minimal value at each SNR point, yielding a composite upper bound. In this way, we obtain a single lower and a single upper bound (denoted as LB, respectively UB). The (semi)numerical results indicate that the newly obtained bounds become close to each other (i.e. tight) in the low-SNR, respectively high-SNR regime, leaving a small gap in the intermediate-SNR regime.
\begin{figure}[h]
	\centering
	\resizebox{0.75\columnwidth}{!}{\includegraphics{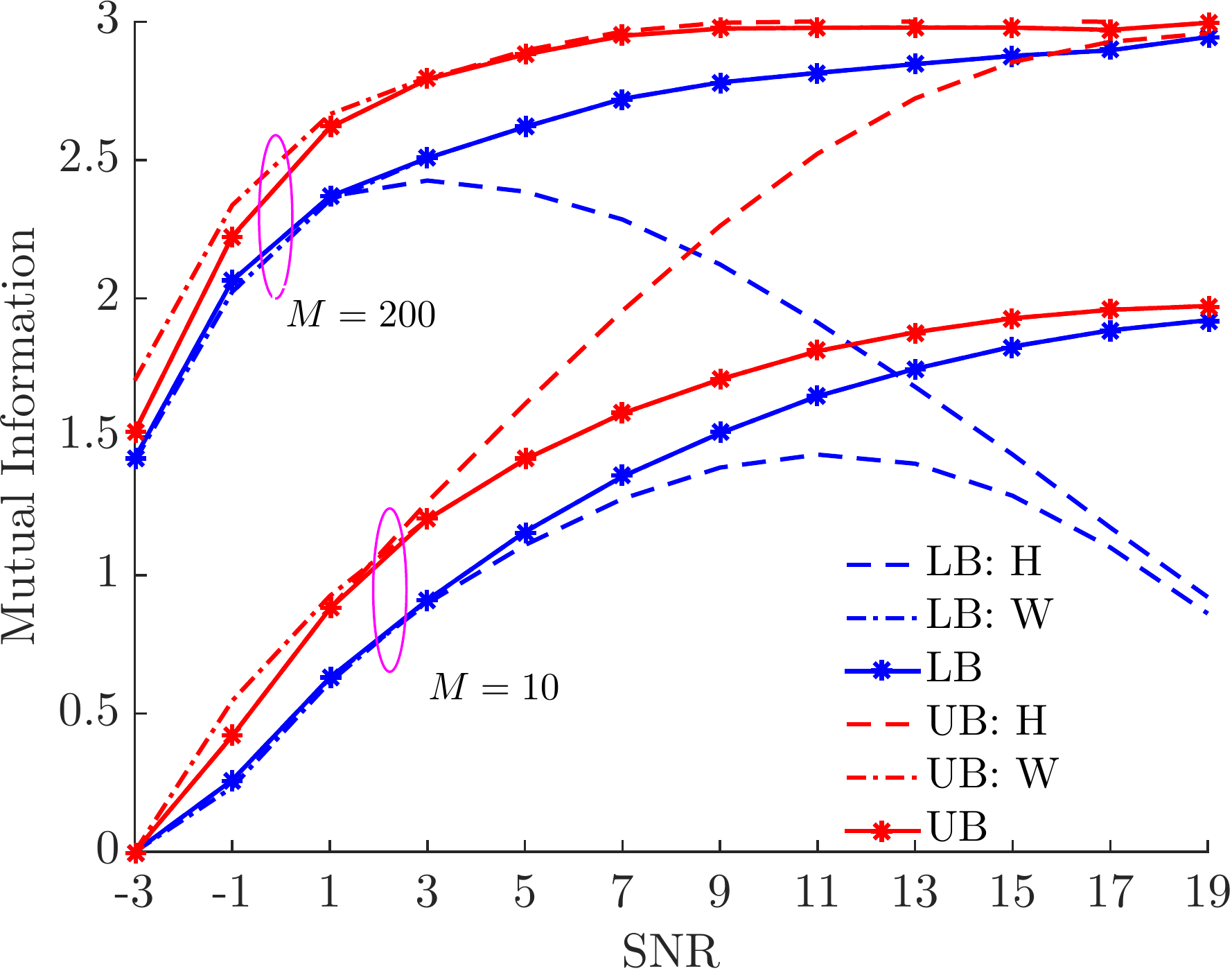}}
	\caption{Mutual information bounds for PAM-ED: $P = 8$.}
	\label{CapaSNRmod8}    
\end{figure}
\noindent Fig. \ref{CapaSNRmod8} illustrates the impact of the SNR and the number of antennas $M$ on the mutual information of a $8$-PAM-ED system. We observe that: $1)$ the mutual information increases significantly with the increase of $M$;  
$2)$ as expected, the lower bound in \eqref{eq:Lower_Bound_1} and the upper bound in \eqref{eq:Upper_Bound_2} represent tight approximations on the mutual information in the high-SNR regime. Meanwhile, the lower bound in \eqref{eq:Lower_Bound_2} and the upper bound in \eqref{eq:Upper_Bound_1} are tight in the low-SNR regime; $3)$ the (composite) lower bound (LB) and the (composite) upper bound (UB) are tight in the low-SNR, respectively high-SNR regime; 4) as SNR increases, UB and LB saturate to a point defined by $M$ and for a sufficiently large $M$ they reach the maximum information rate $\log_2 (|\mathcal{X}|)$. 

\begin{figure}
	\centering
	\resizebox{0.75\columnwidth}{!}{\includegraphics{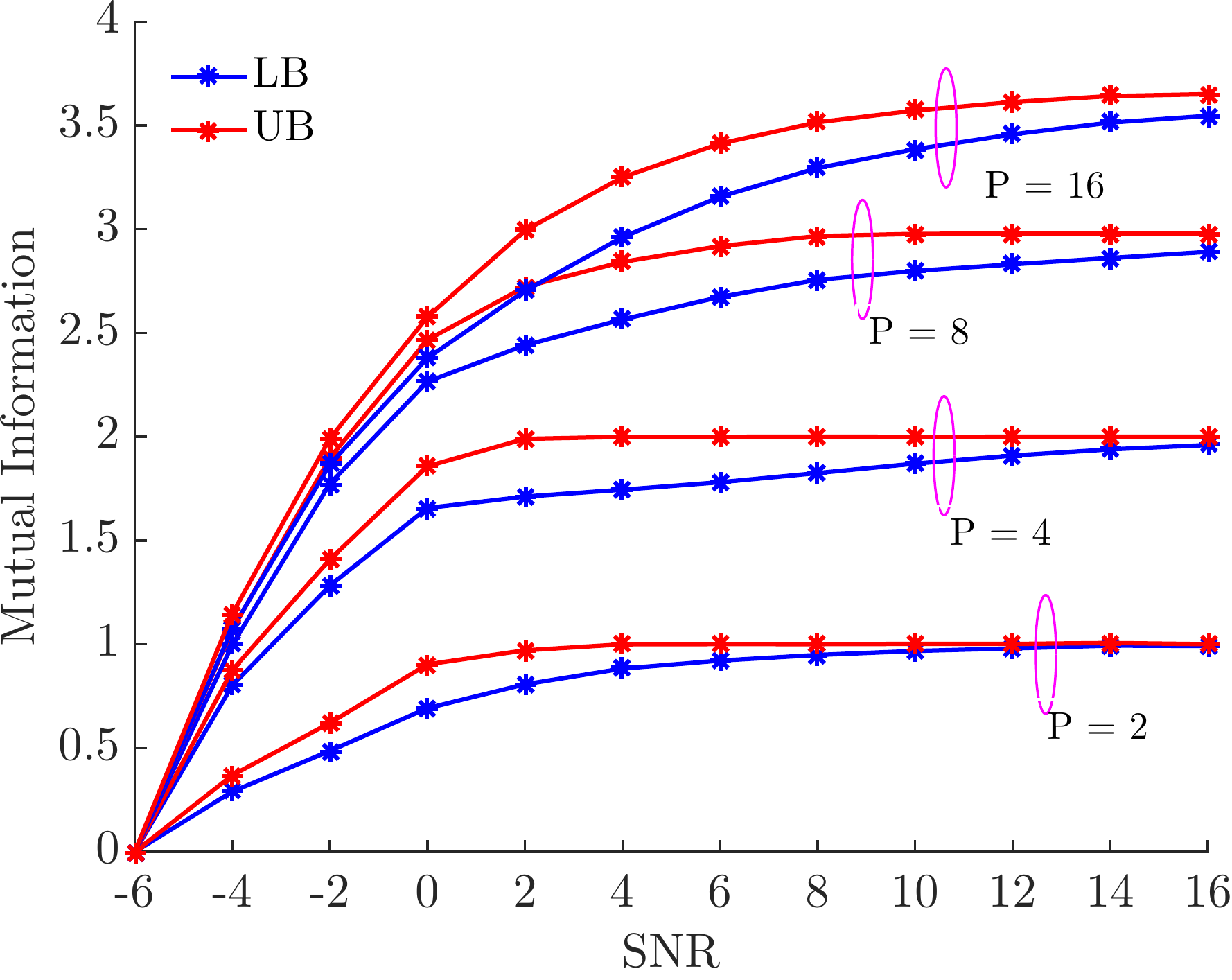}}
	\caption{Mutual information for different modulation order: $M = 200$.}
	\label{CapaAchievableMod100}    
\end{figure}

Fig. \ref{CapaAchievableMod100} reports the impact of the modulation order on the obtained bounds. It also provides insights about the choice of the alphabet size as a function of the SNR and $M$. For example, at SNR = $\SI{6}{\dB}$, it is justified to choose $8$-PAM over $16$-PAM as it achieves (almost) $3$ bits/channel use (the maximum with 8-symbol alphabet), while $16$-PAM achieves a smaller fraction of the maximal achievable rate ($3.5$ out of $4$ bits/channel use).
The observations in Fig. \ref{CapaAchievableMod100} serve as heuristics for the choice of a modulation scheme where the cardinality of the constellation is adopted to the SNR and $M$. A way to select the appropriate constellation size is to define an acceptable performance loss defined as the ratio between the achievable rate (as obtained from UB and LB) and the maximal rate $\log_2 (\vert \mathcal{X}\vert)$. 
\begin{figure}
	\centering
	\resizebox{0.75\columnwidth}{!}{\includegraphics{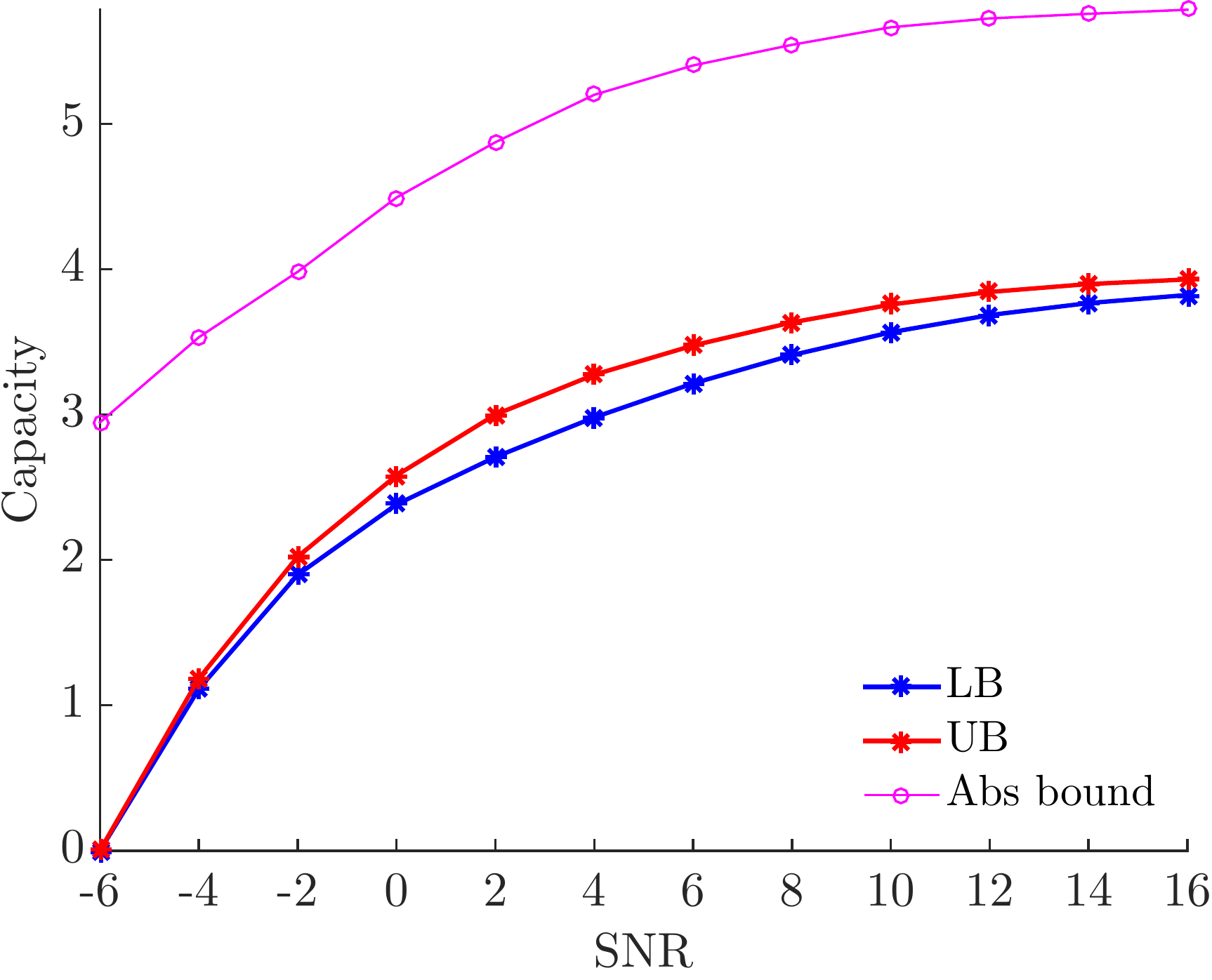}}
	\caption{Mutual information with adaptive modulation:  $ M = 200$.}
	\label{CapaSNR200}    
\end{figure}
In Fig. \ref{CapaSNR200} the achievable rate of this adaptive technique is compared against the capacity of the memoryless Rayleigh fading SIMO channel reported in \cite{MayukhSengupta2000} which provides an absolute bound for all schemes (including ED), under the same channel model. The comparison reveals that PAM-ED seems to follow the $\log M$ scaling behavior of the SIMO capacity. However, there is a gap (albeit SNR-independent) between the achievable rate of the scheme and the SIMO capacity reported in \cite{MayukhSengupta2000}. This behaviour, we think, may be attributed mainly to two effects. First, while ED (implemented by averaging over all antennas) benefits from the ``hardening" effects of both channel and noise, it does not fully exploit the multiple receive antennas. Second, a fraction of the performance loss may be attributed to the suboptimality of PAM. Nevertheless, due to its robustness and the low implementation complexity, ED is definitely justified.
\vspace{-7pt}
\section{Conclusions}
\label{Dsection:conclusion}
In mm-wave systems with a large number of antennas, PAM-ED schemes potentially offer a low-complexity, robust solution to deal with challenges in CSI acquisition. We have investigated the impact of the number of antennas and SNR on the achievable information rate and provided bounds on the performance. We have shown that, when applied in an adaptive fashion, the performance of PAM-ED scales accordingly with the number of antennas and the SNR. The performance, together with the robustness and the simple implementation, justify the proposed use of ED in practical mm-wave systems. It remains for future work to evaluate the performance of ED when implemented at each receive antenna separately. Although more complex in the implementation, we suspect that this implementation may bridge the gap to the SIMO capacity.   
\vspace{-10pt}
\section*{Acknowledgment}
The work of L. Jing and E. de Carvalho has been in part supported by the Danish Council for Independent Research (Det Frie Forskningsr\aa d), DFF $133500273$. The work of Z. Utkovski has been in part supported by the
German Research Society (DFG) via the project Li 659/13-1. The work of
P. Popovski has been in part supported by the European
Research Council (ERC Consolidator Grant nr. 648382 WILLOW). 
\bibliographystyle{IEEEtran}
\vspace{-7pt}
\bibliography{IEEEabrv,AAURef}
\end{document}